# GaMnAs grown on (311) GaAs substrates: modified Mn incorporation and new magnetic anisotropies

K. Y. Wang[1], K.W. Edmonds[1], L. X. Zhao[1], M. Sawicki[2], R. P. Campion[1], B.L. Gallagher[1], C.T. Foxon[1]

1School of Physics and Astronomy, University of Nottingham, Nottingham NG7 2RD, UK

2Institute of Physics, Polish Academy of Sciences, and ERATO Semiconductor Spintronics Project, PL 02-668 Warszawa, Poland

**Abstract**

We report the results of a detailed study of the structural, magnetic and magnetotransport properties of as-grown and annealed $Ga_{0.91}Mn_{0.09}As$ thin films grown on (311)A and (311)B GaAs substrates. The high Curie temperature and hole density of the (311)B material are comparable to those of GaMnAs grown on (001) GaAs under the same growth conditions, while they are much lower for the (311)A material. We find evidence that Mn incorporation is more efficient for (311)B than for (001) and significantly less efficient for (311)A which is consistent with the bonding on these surfaces. This indicates that growth on (311)B may be a route to increased Curie temperatures in GaMnAs. A biaxial magnetic anisotropy is observed for the (311) material with easy axes along the [010] and [001] out-of-plane directions. An additional uniaxial in-plane anisotropy is also observed with the easy axis along $[01\bar{1}]$ for the (311)A material, and along $[\bar{2}33]$ for the (311)B material. This new observation may be of importance for the resolution of the outstanding problem of the origin of uniaxial anisotropy in (001) GaMnAs.

## Introduction

The ferromagnetic semiconductor GaMnAs has attracted considerable interest, mainly because of the prospect of developing devices which combine information processing and storage functionalities in one material [1,2]. When Mn atoms occupy the Ga site they act as acceptors. The Curie temperature $T_C$ of GaMnAs is strongly dependent on both the effective Mn concentration $x_{eff}$ and the hole density p, varying roughly as $x_{eff}p^{1/3}$ [3]. GaMnAs films grown on GaAs (001) substrates by low-temperature molecular beam epitaxy are under compressive strain, which for sufficiently high hole density leads to an in-plane magnetic easy axes [4]. The magnetic properties of GaMnAs as-grown samples can be phenomenologically described by a single domain model considering co-existing in-plane biaxial and uniaxial anisotropies [5-7].

Growth of GaMnAs on higher index substrates has attracted much less attention. This is surprising given that this may offer a new degree of freedom for engineering magnetic anisotropies and maximising incorporation of magnetic impurities. There are reports of GaMnAs grown on (411)A GaAs [8], and very recently, on (311)A GaAs [9]. In both cases, magnetotransport measurements gave evidence for magnetic easy axes aligned partially along the growth direction, in the (001) plane, although the film quality was rather low compared to GaMnAs grown on (001) GaAs. In this paper we report the results of a comparative study of the structural, magnetic and magnetotransport properties of as-grown and annealed $Ga_{0.91}Mn_{0.09}As$ thin films grown on (311)A and (311)B GaAs substrates. We examine in detail the resulting magnetic anisotropies, and contrast these to the widely-reported behaviour of GaMnAs on (001) GaAs. We also investigate the differing efficiencies of Mn incorporation on these surfaces and the effect of this on the Curie temperatures obtained.

**Experimental**

50nm thick $Ga_{0.91}Mn_{0.09}As$ films were grown on (311)A and (311)B substrates by low temperature molecular beam epitaxy under identical growth conditions to those used for (001) GaMnAs. The growth [10], structural [11], magnetic [12] and magnetotransport [13] properties of the (001) GaMnAs layers are described in detail elsewhere. A 100nm thick high-temperature GaAs layer, followed by a 50nm thick low-temperature GaAs layer, were deposited prior to growth of the GaMnAs films. The nominal Mn concentration of 9% is obtained from the Ga/Mn flux ratio, as calibrated by secondary ion mass spectrometry measurements on 1μm thick films grown on (001)GaAs. We note however that it may be substantially different for the (311) materials for which the efficiency of Mn incorporation may be significantly different from that for (001). Post-growth annealing was performed in air at 190°C for around 100 hours, which is an established procedure for increasing the $T_C$ of thin GaMnAs films [14,15]. The structural properties were measured by high resolution X-ray diffraction. To obtain accurate relaxed lattice parameters, the ω/2θ curves obtained are compared with simulations using Philips X'Pert epitaxy software package of the PANalytical X'Pert XRD system. Standard hall bar structures were photolithographically fabricated for transport measurements, with the current direction along the [$\bar{2}$33] axis. Magnetotransport measurements were performed, using low frequency ac techniques, in a cryo-magnetic system, in which the magnetic field can be rotated *in-situ* along any direction relative to the sample axes. The magnetic properties of the layers were measured using a SQUID magnetometer, which has been custom designed to give very low trapped fields in the superconducting coil (<0.1 Oe) after a field sweep up to 1000 Oe.

**Results and discussion:**

**i) Structure**

Figure 1 shows experimental and simulated X-ray diffraction 2θ/ω plots for the as-grown, (311)A and B GaMnAs samples. Features corresponding to the GaAs substrate, the low-temperature GaAs buffer layer, and the GaMnAs layer are marked. The lattice constant in the LT-GaAs buffer layer is slightly expanded compared to the substrate, probably due to $As_{Ga}$ defects. The GaMnAs peak is much broader for (311)A compared to (311)B indicating poorer structural quality, and it is at higher 2θ/ω. The width of the GaMnAs peak for the (311)B film is similar to that of the (001) film, indicating comparable structural quality. The lattice constant obtained for (311)B samples is significantly larger than that for (311)A samples. Since the lattice constant in GaMnAs increases with increasing Mn concentration [16], this indicates that less Mn has been incorporated in the (311)A case. As shown in table I, the strain $\Delta a/a$ is the largest for 311B in these three samples, which indicates the most efficient Mn incorporation for 311B.

**ii) Magnetotransport**

The resistivity (ρ) as a function of temperature (T) for the as-grown and post-annealed (311) GaMnAs films is shown in Fig.2, together with data for a 50nm (001) GaMnAs film grown under the same conditions. For (311B), the resistivity exhibits a clear peak at temperature $T_ρ$ corresponding roughly to the onset of ferromagnetic ordering, as is typically observed in GaMnAs films on the metallic side of the metal-insulator transition (MIT) [16,17]. On annealing, the resistivity decreases while $T_ρ$ moves to higher temperatures. Both the resistivity and $T_\rho$ are comparable to those obtained for the (001) sample. In contrast, the resistivity of the (311)A film is much higher. The film is on the insulator side of the MIT before annealing, and close to the transition after annealing. Room temperature Hall measurements show that all the samples are *p*-type. Low temperature high field Hall measurements [17,18] give the hole densities listed in table I for the (001) and (311)B samples. At low temperatures, the resistance is too high to accurately study the magnetotransport properties of (311)A samples. However, it can be clearly seen from table I that the electrical and magnetic properties are better for the (311)B film than for the (311)A film, and that these properties improve on annealing of both films.

For the (311)B sample, Hall measurements show a clear, albeit complicated, Hall response. The measured Hall resistance, $R_{Hall}$, versus magnetic field applied along the [311] and [010]/[001] directions are shown in Fig. 3a for the as-grown (311)B sample. The 3D schematic diagram for the 311 material is shown in the inset of Fig.3a. We note here, that one can unambiguously distinguish the $(\bar{2}33)$ and $(01\bar{1})$ planes. Looking from the top, the upright plane is $(01\bar{1})$ plane and the cutting edge plane is $(\bar{2}33)$ plane. [010]/[001] directions are tilted 17.5° out of the film plane. Along [010]/[001], $R_{Hall}$ has a square loop which is offset from zero, followed by a slow decrease; while for the [311] directions the loop is even more complicated, showing a crossover of increasing and decreasing field sweeps. The measured $R_{Hall}$ is a superposition of two distinct contributions. Firstly the anomalous Hall contribution which is given by $R_A = \alpha_A \rho^\gamma M_z$, where $\alpha_A$ is the anomalous Hall coefficient, and $M_z$ is the perpendicular component of the magnetisation. For GaMnAs in the metallic regime $\gamma \sim 2$ [18]. The second contribution is from the so-called Planar Hall effect, which arises due to a difference in conductivity for current parallel and perpendicular to the magnetisation direction. This is approximately given by $R_P = \alpha_P \sin(2\theta)$, where $\theta$ is the angle between magnetisation and current directions. The large Planar Hall contribution observed when the field is applied along [311] indicates that $\theta$ varies during the magnetisation reversal.

The Planar Hall effect is not a true Hall effect, i.e. is not antisymmetric in the magnetic field, so the anomalous Hall resistance can be extracted from the measured $R_{Hall}$ using the expressions:

$R_A(\downarrow) = [R_{Hall}(\downarrow,+) - R_{Hall}(\uparrow,-)]/2$ and $\quad R_A(\uparrow) = [R_{Hall}(\uparrow,+) - R_{Hall}(\downarrow,-)]/2$,

where ↑, ↓, +, - represent increasing, decreasing, positive and negative external magnetic fields, respectively. The extracted $R_A$ for field along the [311] and [010] / [001] directions is shown in Fig. 3b. For both directions, a sharp switch between ±60Ω is observed at low fields. At higher fields, $R_A$ (and thus $M_z$) increases slowly and reversibly for the [311] direction, which is characteristic of rotation away from an easy magnetic axis, while for the [010]/[001] directions $R_A$ remains almost constant above the switching field of ~300 Oe. This indicates that the magnetisation is saturated along this direction (note that the small decrease in $R_{Hall}$ at higher fields is due to its dependence on $\rho$, which shows an isotropic negative magnetoresistance as is commonly observed in (001) GaMnAs [16]. Therefore, the Hall resistance data indicate that the (311)B film has easy axes along the [010] and [001] directions, which are tilted out-of-plane. Square Hall traces are also observed for the annealed (311)B film for field parallel to the [010]/[001] directions, as shown in the inset of Fig. 3b. Similar anisotropies have been reported previously for GaMnAs grown on (411)A GaAs [8], and have been ascribed to the strain-induced anisotropy of the GaMnAs valence band.

**iii) Magnetometry**

SQUID hysteresis loops for the as-grown and annealed (311)A and B samples at 5K are shown in Fig. 4, for magnetic fields applied along the orthogonal $[01\bar{1}]$, $[\bar{2}33]$, and [311] axes (it proved impossible to perform reliable magnetometry measurements with the applied field along the <100> directions in the SQUID magnetometer). The diamagnetic contribution from the substrate has been subtracted. The measured magnetization is normalised to the nominal Mn concentration. The magnetization is larger for (311)B than for (311)A, again indicating that a larger amount of the incident Mn is incorporated substitutionally in the former case. The magnetisation increases on annealing, as is typically observed in (001) GaMnAs films. As shown in table I, for the annealed (311)B film the measured magnetic moment per Mn is (3.9±0.3) $\mu_B$, which is even much larger than the value of (2.6±0.3) $\mu_B$ observed for the (001) material. This is again consistent with a larger fraction of the nominal Mn concentration of 9% being incorporated substitutionally in the (311)B film than in (001) film. Since the surface As atoms are twofold coordinated on the (311)A surface and threefold coordinated on the (311)B surface, this indicates that Mn is incorporated more readily at a single-dangling bond site. A similar situation is also observed in Si-doping of (311)GaAs, where incorporation on As and Ga sites is favoured for growth on A and B surfaces respectively. This has been ascribed to a stronger bonding to the substrate of adatoms at double-dangling bond sites, so that breaking of the weaker single-dangling bonds is the limiting impurity incorporation process [19].

For all four (311) samples and for all three orientations, significant hysteresis is observed, with larger remnant magnetization $M_R$ along the in-plane orientations than along the perpendicular [311] direction. The angles between the [$01\bar{1}$], [$\bar{2}33$], and [311] axes and the [010]/[001] axes are respectively 45°, 50.2°, and 72.5°. Therefore, if the measured $M_R$ represents the projection of the magnetization which is oriented along the easy [010]/[001] axes (as suggested by the Hall measurements on the (311)B film), we expect the $M_R$ along [$01\bar{1}$], [$\bar{2}33$], and [311] to have the ratio cos(45°): cos(50.2°): cos(72.5°). This ratio is observed almost exactly for the as-grown (311)A sample, and approximately for the annealed (311)A and as-grown (311)B samples, although with a larger in-plane moment in the latter two. However, for the annealed (311)B sample, the $M_R$ along [$\bar{2}33$] is by far the largest, with a pronounced anisotropy between this and the [$01\bar{1}$] direction. These measurements indicate that the films also possess a uniaxial magnetic anisotropy which favours in-plane orientation of the magnetization, and the ratio of the strength of the uniaxial to biaxial anisotropy increases on going from as-grown (311)A, to annealed (311)A, to as-grown (311)B, to annealed (311)B, i.e. following the order of increasing $T_C$.

To obtain more detailed information about the magnetic anisotropy and its temperature-dependence in the films, the temperature-dependence of the remnant magnetization is studied. First, the sample is cooled through $T_C$ down to 5K under an external field of 1000Oe applied along the measurement axis. Then, the field is removed, and the projection of the remnant magnetization along the measurement axis is recorded as a function of increasing temperature. Since the trapped fields in the superconducting coil are very low, the sample magnetization rotates to the nearest magnetic easy direction set only by the torques exerted by internal anisotropy fields. These measurements were performed sequentially for the orthogonal [$01\bar{1}$], [$\bar{2}33$], and [311] orientations.

The temperature dependence of the $M_R$ for the as-grown and annealed (311)A and (311)B films are shown in Fig. 5. The Curie temperatures obtained from this procedure are in agreement with those obtained from transport measurements, shown in Table I. The measured remnant magnetization projections indicate the presence of both in-plane uniaxial and [001]/[010] biaxial anisotropies, consistent with the hysteresis measurements. With increasing temperature, the uniaxial anisotropy becomes dominant. The uniaxial easy axis is along [$01\bar{1}$] in the (311)A case and along [$\bar{2}33$] in the (311)B case, both before and after annealing.

It is useful to compare the magnetic anisotropy of the (311) films to previously reported behaviour for (001) GaMnAs under compressive strain [5,6,7,12,13,20,21]. In such films, biaxial anisotropy with in-plane [100] and [010] easy axes typically dominates at low temperatures, giving way to a uniaxial anisotropy between in-plane [110] and [1-10] directions at high temperatures. Low temperature annealing tends to enhance the in-plane uniaxial anisotropy relative to the biaxial [20], and causes a rotation of the uniaxial easy axis from the [1-10] direction to the [110] [12]. The origin of the symmetry breaking between [110] and [1-10] in (001) GaMnAs is presently unknown.

While a uniaxial anisotropy is expected in the present case since the $[01\bar{1}]$ and $[\bar{2}33]$ directions are not symmetric, the different orientation of the uniaxial easy axis is somewhat surprising since the structures of (311)A and (311)B differ only at the surface and interface layers. It has been suggested for (001) GaMnAs [21] that the uniaxial anisotropy results from an anisotropic incorporation of Mn during growth. However, for the 311A and 311B surfaces, similar (8×1) surface reconstruction along $[01\bar{1}]$ direction has been observed for (311)A and (311)B GaAs samples [22,23] and the Ga bonds on both surfaces align along the $[0\bar{1}1]$ direction. Therefore, an anisotropic incorporation is unlikely to be responsible for the difference in the uniaxial anisotropy between (311)A and (311)B. We therefore attribute the uniaxial anisotropy rotation to the higher incorporation on the B surface, leading to a higher hole density, since the direction of the uniaxial easy axis has been shown to be hole-density dependent in the case of (001) GaMnAs. However, we cannot rule out that the higher concentration of local moments also plays a role.

The temperature-dependent magnetization under a 1000 Oe external field applied along the $[\bar{2}33]$ direction for the annealed (311)B sample is also shown in figure 5. After subtracting the diamagnetic background from the substrate, this is almost identical to the remnant measurement. This indicates that, after field-cooling to align the magnetization along the $[\bar{2}33]$ direction, the $M_R$ remains fixed along this direction on warming. Differences are only observed close to $T_C$, due to the paramagnetic response of the GaMnAs layer under magnetic field, and at very low temperatures, where comparison of the $M_R$ traces for the three orientations appears to suggest that the easy axis rotates away from $[\bar{2}33]$ and towards the [010]/[001] directions. However, applying a field along the $[01\bar{1}]$ or [311] directions results in a net $M_R$ along these directions, indicating that the magnetization does not rotate into the $[\bar{2}33]$ easy axis on sweeping the external field to zero. From this we conclude that the $M_R$ is partially oriented along the [010] and [001] axes after applying a field along $[01\bar{1}]$ or [311]. Assuming that the measured $M_R$ along these two directions corresponds to the projection of the [010] and [001] magnetization, while the magnetization which is not detected is aligned with the $[\bar{2}33]$ direction, we obtain the components of the magnetization along the [010]/[001] and $[\bar{2}33]$ directions after field-cooling along $[01\bar{1}]$ or [311] as shown in Fig. 6. Quite different results are obtained for $[01\bar{1}]$ and [311], with the latter giving rise to a larger magnetization along the [010]/[001] directions. This demonstrates that the magnetic state may be strongly history-dependent in (311) GaMnAs, with large energy barriers separating uniaxial in-plane and biaxial out-of-plane easy axes.

**Summary**

We have studied the structural, magnetotransport and magnetic properties of $Ga_{0.91}Mn_{0.09}As$ thin films grown on (311)A and (311)B GaAs substrates. The larger lattice constant, Curie temperature, conductivity and saturation magnetization observed for the (311)B sample are all evidence that a higher fraction of incident Mn is incorporated substitutionally at Ga sites for growth on (311)B than on (311)A. For both (311)A and (311)B, a factor of two increase in $T_C$ is obtained on annealing, indicating that Mn is also incorporated on interstitial sites in both cases [24]. Despite evidence for higher Mn incorporation for (311)B than for (001) the materials show comparable $T_C$, hole density and conductivity suggesting that the concentration of compensating defects such as $As_{Ga}$ may also be larger. Therefore, with further optimization growth on (311)B may be a route to higher temperature ferromagnetism in GaMnAs.

The (311) GaMnAs films show a complex and interesting magnetic anisotropy, with competition between out of plane biaxial [010]/[001] easy axes, and an in-plane uniaxial easy axis. With increasing temperature, the in-plane easy axis becomes dominant. Low temperature annealing strongly enhances the in-plane uniaxial anisotropy relative to the biaxial. Similar behaviour has also been observed in (001) material. Moreover, the in-plane easy axis is along the $[01\bar{1}]$ direction for the (311)A sample and along the $[\bar{2}33]$ direction for the (311)B sample. While the origin of the easy axis rotation is not presently known, this observation may hold important clues regarding the origin of the in-plane uniaxial anisotropy observed in (001) GaMnAs.

**Acknowledgements**

The project was funded by EPSRC (GR/S81407/01) and EU FENIKS (G5RD-CT-2001-0535). The authors gratefully acknowledge useful discussions with T. Dietl, C. Staddon, and M. Henini. We also thank Jaz Chauhan and Dave Taylor for processing of the Hall bars.

# Figure captions

Figure 1. ω/2θ plots for (Ga,Mn)As grown on (311)A and (311)B GaAs. Features corresponding to the GaAs substrate (filled up triangle) LT-GaAs buffer (filled down triangle) and GaMnAs layer (open down triangle) are marked. The gray curves are experimental data and the black curves are simulations.

Figure 2. Resistivity versus temperature for as-grown and annealed (Ga,Mn)As films grown under similar conditions on (001) (circles), (311)A (squares) and (311)B (triangles) GaAs substrates, where the peak temperature $T_\rho$ are marked with arrows. The solid symbols are for as grown samples the open symbols are for annealed samples.

Figure 3 (a). Hall resistance, including planar Hall contribution, for as-grown (Ga,Mn)As on (311)B GaAs at 4K, for magnetic field applied along the [311] (squares) and [010] / [001] (circles) directions. Inset of (a) the 3D schematic diagram for 311 material: (b) extracted anomalous Hall resistance for the data in (a); inset of (b). Hall resistance for field along the [010] / [001] direction at 4K for annealed (Ga,Mn)As film on (311)B GaAs.

Figure 4. Magnetisation hysteresis loops at 5K for three orthogonal directions [$01\bar{1}$] (circles), [$\bar{2}33$] (squares) and [311] (triangles), for (a) as-grown (Ga,Mn)As on (311)A GaAs; (b) annealed (Ga,Mn)As on (311)A GaAs; (c) as-grown (Ga,Mn)As on (311)B GaAs (measurement along [$01\bar{1}$] not shown); (d) annealed (Ga,Mn)As on (311)B GaAs. The diamagnetic contribution from the substrate has been subtracted in each case.

Figure 5. Remnant magnetisation along the [$01\bar{1}$] (circles), [$\bar{2}33$] (squares) and [311] (triangles) axes versus increasing temperature, after cooling from above $T_C$ under a 1000 Oe field, for (a) as-grown and annealed (Ga,Mn)As on (311)A GaAs and (b) as-grown and annealed (Ga,Mn)As on (311)B GaAs. Measurements on as-grown and annealed films are denoted by full and open symbols respectively. The solid line in (b) is the magnetisation along [$\bar{2}33$] under a 1000 Oe field, where the diamagnetic contribution from the substrate has been subtracted.

Figure 6. Estimated (see text) magnetisation oriented along [010]/[001] and [$\bar{2}33$] orientations versus increasing temperature after cooling under an external field applied along (a) [$01\bar{1}$] and (b) [311] axes, for the annealed (Ga,Mn)As film on (311)B GaAs.

**Table I** the strain $\Delta a / a$, hole density p, Curie temperature $T_C$ and saturation magnetization $M_S$ at 5 K for 50 nm thick $Ga_{0.91}Mn_{0.09}As$ films grown on (001), (311)A and (311)B GaAs substrates. Here $a$ is relaxed lattice constant of GaMnAs epilayer and $\Delta a = a - a_0$, where $a_0$ is lattice constant of GaAs substrate.

**Figure 1 K. Y. Wang et al.**

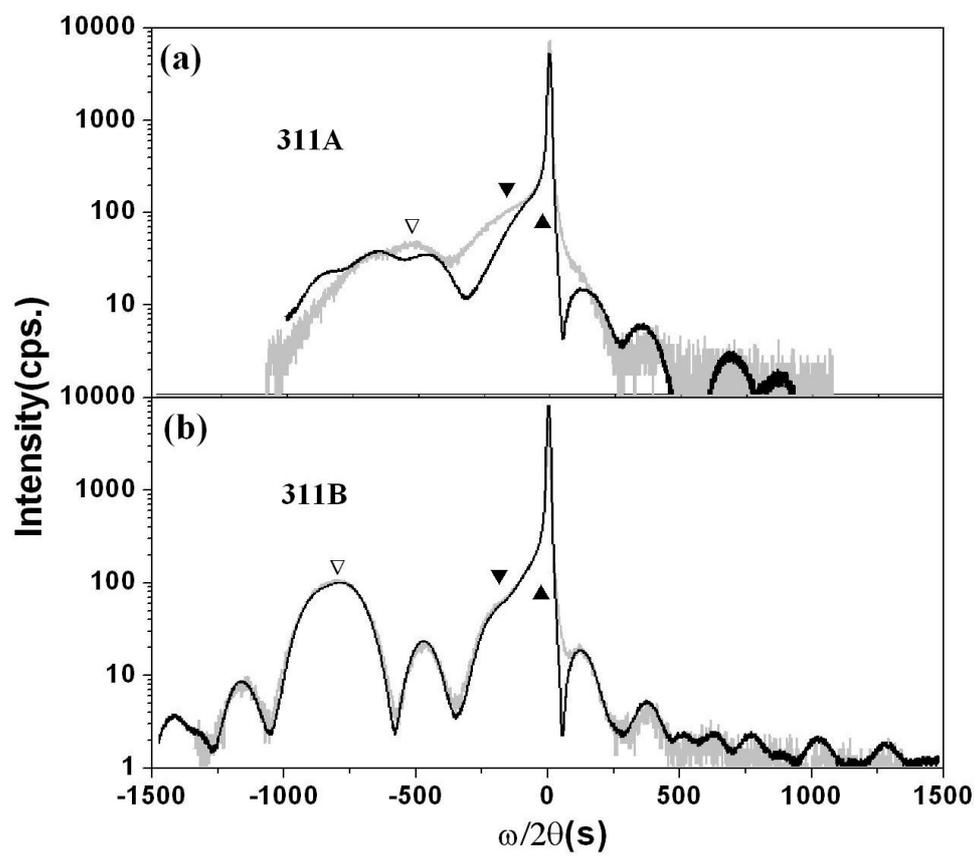

**Figure 2  K. Y. Wang et al.**

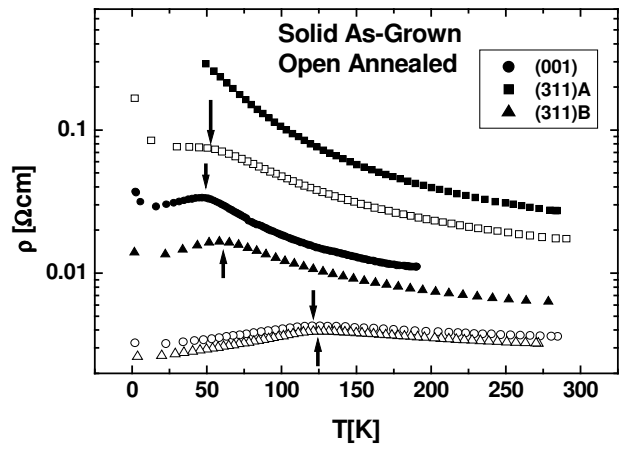

**Fig.3 K. Y. Wang et al.**

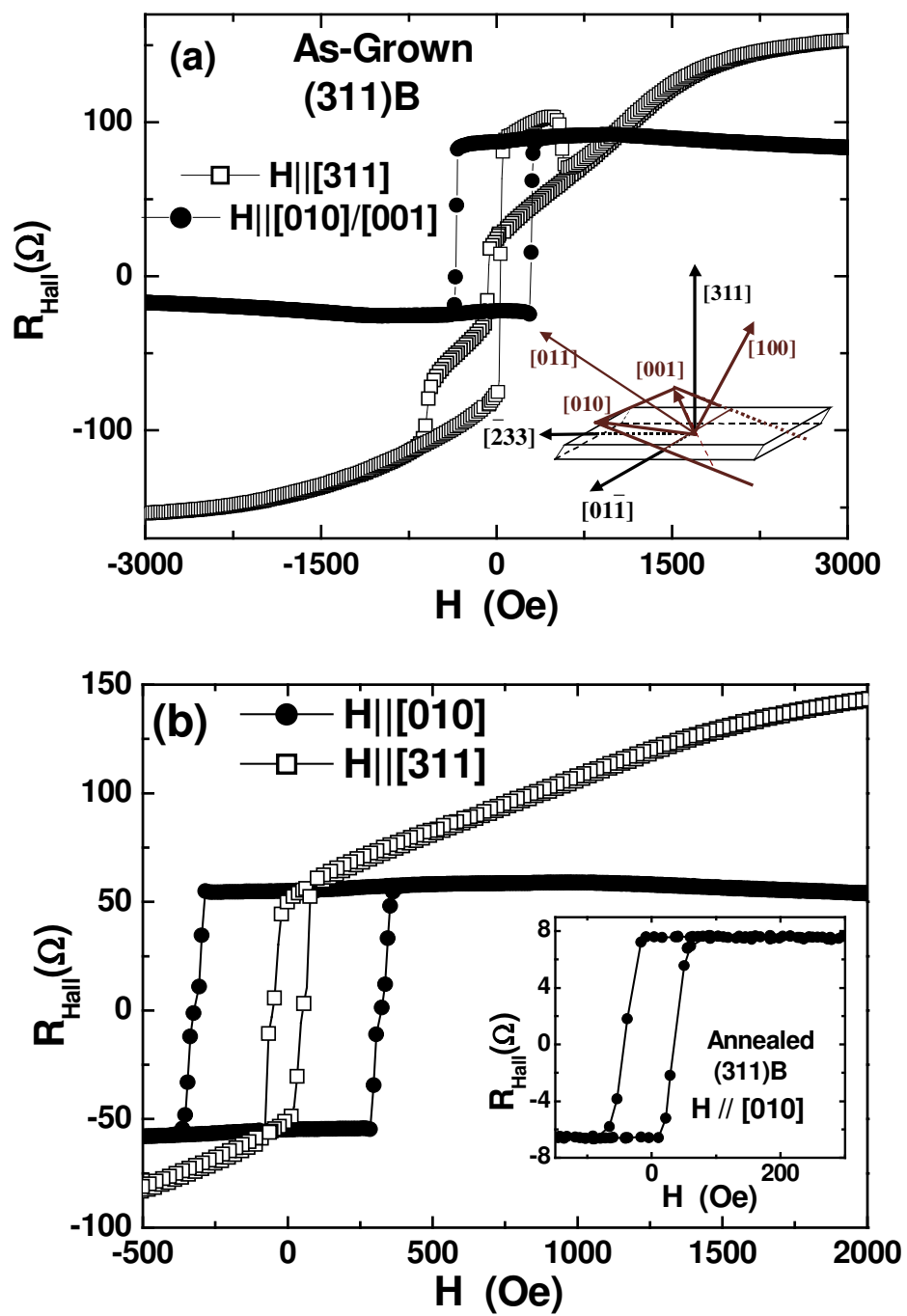

**Fig.4 K. Y. Wang et al.**

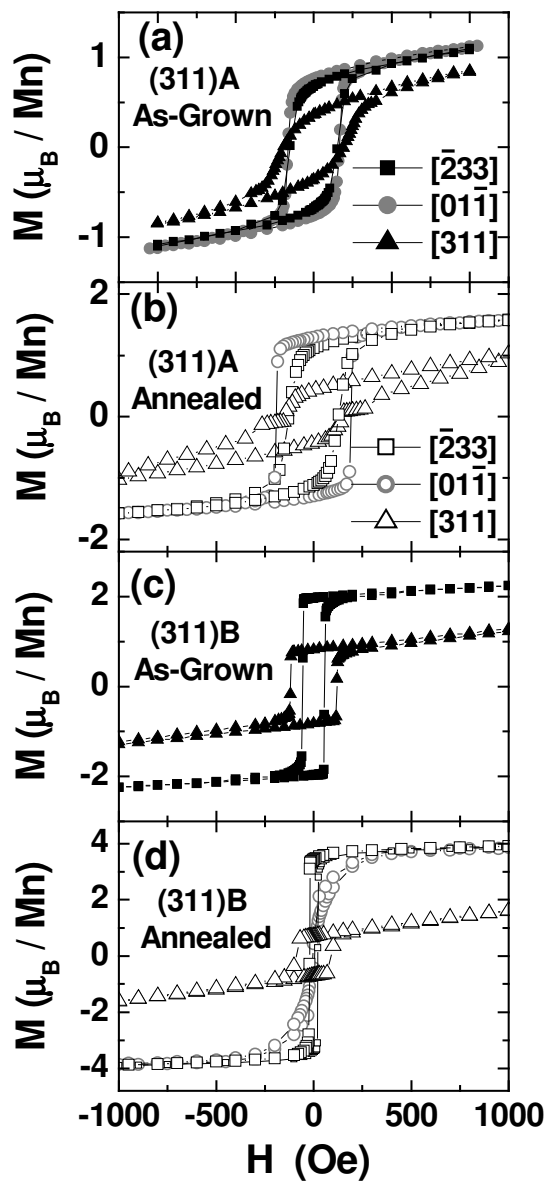

**Fig.5 K. Y. Wang et al.**

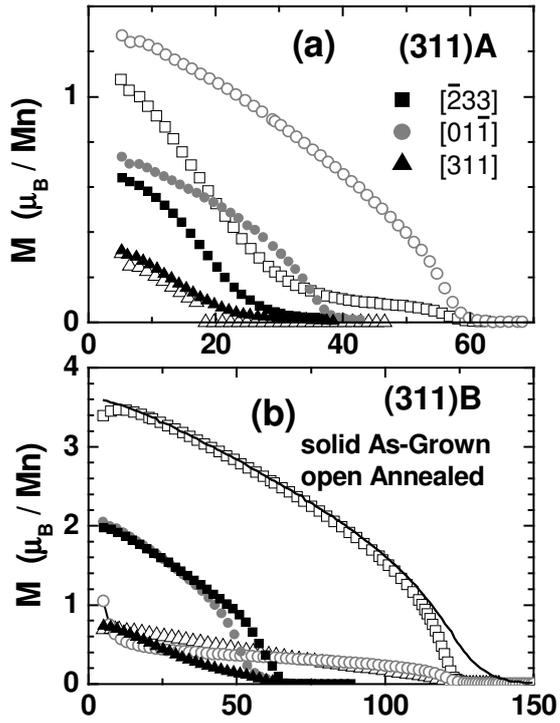

**Fig.6 K. Y. Wang et al.**

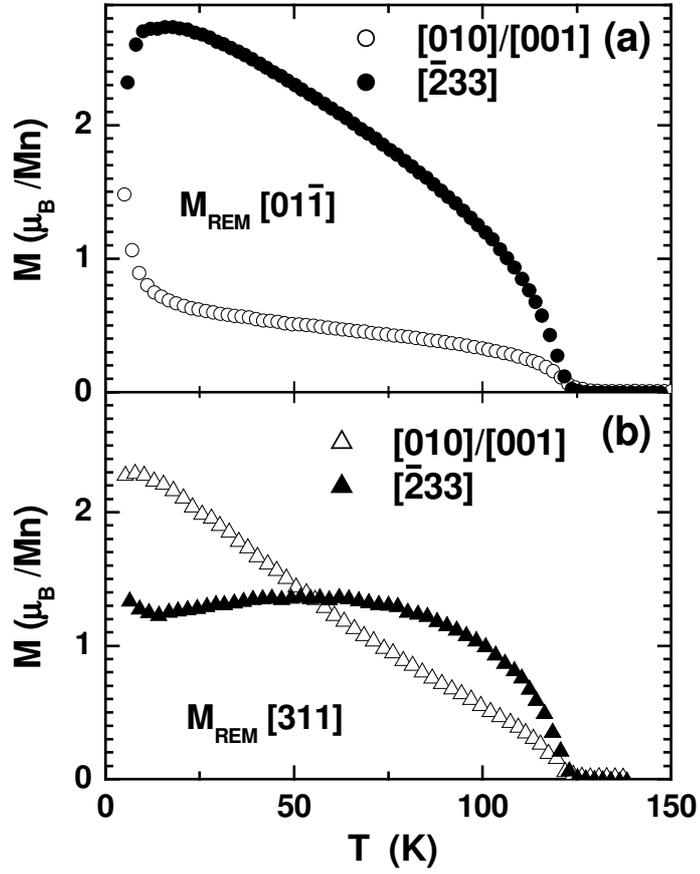

**Table I K. Y. Wang et al.**

|  | **(001)** | **(001)ann** | **(311)A** | **(311)A$_{ann}$** | **(311)B** | **(311)B$_{ann}$** |
|---|---|---|---|---|---|---|
| $\Delta a/a$ (×$10^{-3}$) | 3.65 | ~ | 3.1 | ~ | 4.5 | ~ |
| p±0.3 (×$10^{20}$cm$^{-3}$) | 0.9 | 5.5 | ~ | 0.31 | 0.92 | 5.4 |
| $T_C$±2(K) | 56 | 133 | 38 | 62 | 62 | 125 |
| $M_S$±0.3($\mu_B$/Mn) | 2.1 | 2.6 | 1.2 | 1.7 | 2.2 | 3.9 |